\documentstyle[floats,epsf,aps]{revtex}

\begin{document}
\bigskip
\newcommand{\beq}{\begin{equation}}
\newcommand{\eeq}{\end{equation}}
\renewcommand{\textit}[1]{{\large #1}}
\large
\title{
{\bf On the Flow of Kaons Produced in Relativistic Heavy Ion Collisions} } 
\author{ {\bf C. David, C. Hartnack and J. Aichelin} 
\\
{\normalsize
SUBATECH}\\
{\normalsize
 Universit\'e de Nantes, EMN, IN2P3/CNRS}\\
{\normalsize    4, Rue Alfred Kastler, BP 20722, 44307 Nantes Cedex 3, France }}

\maketitle

\

{\Large
\begin{abstract} \large
{\bf Abstract:} We investigate the different contributions to the in-plane 
flow of $K^+$ mesons observed recently by the FOPI collaboration in the reaction
Ni (1.93 A GeV) + Ni. Due to the kinematics of the three body phase space decay
the flow of the kaons produced in baryon-baryon interactions is smaller than 
that of the baryons in the entrance channel. On the contrary, 
in $\pi$ N interactions the flow of the sources and of the kaons are
identical. Therefore the total kaon flow depends on the relative number of 
$\Delta N  \rightarrow K^+$ and $\pi N \rightarrow K^+$ reactions and hence 
on
the lifetime of the $\Delta$, in addition to the already known dependence on
the potential interaction of the kaons with the nuclear environment.

\end{abstract}

\section{introduction}
The production of K mesons in heavy ion collisions is presently one of the
most challenging topics in nuclear physics. At beam energies below or
close to the threshold (in NN collisions $E_{thres} = 1.583$ GeV) we
observe a strong enhancement of the kaon production as compared to the
extrapolation of pp collisions. Detailed investigations have shown that
most of the K's are created in two step processes via an
intermediate $\Delta$ or $\pi$ and are produced at a density well above
the normal nuclear matter density \cite{aik,har}. This triggered the conjecture that 
K's may be of use as a messenger of the high density zone. 

What makes a straight forward analysis complicated are several problems:
\begin{itemize}
\item  The nucleons as well as the K's and $\Lambda's$ interact with each other via
static  and momentum dependent interactions.  Hence particle properties like
the particle mass are
modified. 
\item This modification of the particle properties changes the thresholds of
the relevant strange particle production cross sections $ N N \rightarrow 
\Lambda K N$,  
$ N N \rightarrow \Sigma K N$,  $ N N \rightarrow N N\bar K K $ as well as
of the the rescattering reactions. 
\item  
Most of the K's are produced in a collision in which a nuclear resonance or
its decay product is involved. Hence the lifetime of a resonance in nuclear
matter becomes important. If the life time $\tau$ is short as compared to 
$\lambda(\rho)/<v>$, where $ \lambda$ is the density dependent mean free path
of the nuclear resonance and $<v>$ is its average velocity,
the resonance decays before it encounters another nucleon for creating a kaon.
Then only the decay product, i.e. the $\pi$ can create a kaon. If the opposite is the case,
the dominant production channel will be Resonance + N $\rightarrow 
\Lambda (\Sigma) N K$.
\end{itemize}
It is the purpose of this article 
to investigate the influence of these
in medium properties on the production of K's, especially on the observed
in-plane flow which has been proposed as a signature of the relative
strength of the scalar and vector part of the interaction of the K's
with its hadronic environment \cite{ko7}.

For this purpose we employ simulations of the heavy ion reaction with
the Quantum Molecular Dynamics (QMD) approach. This approach is described
in ref \cite{aic}. In addition we have implemented the kaon producing cross
sections $\pi B\rightarrow YK$ and $BB\rightarrow BYK$ with $B=N,\Delta$ and
$Y=\Lambda,\Sigma$. We have displayed in fig. 1 the important cross section
with a $\Lambda$ in the final state. Similar fits to the (fewer) available data
for the reaction with a $\Sigma$ in the final state have also been made and are
included in the calculation. The energy range of importance is shown in fig. 2
where we display the number of elementary collisions as a function of
$\sqrt{s}$ in the simulation. The relevance of the different channels is
given by the number of collisions times the kaon production cross section.
We see that $\pi N$ ,$\Delta$ N and NN collision contribute all to the
kaon production in a non negligible way. We see as well that the deviations
of our parameterization from data at high $\sqrt{s}$ is not of importance for
the beam energy considered here.

After being produced the K's move in a potential
created by the nuclear environment which has the form \cite{scha}:

\begin{figure}[h]
\epsfxsize=11.cm
$$
\epsfbox{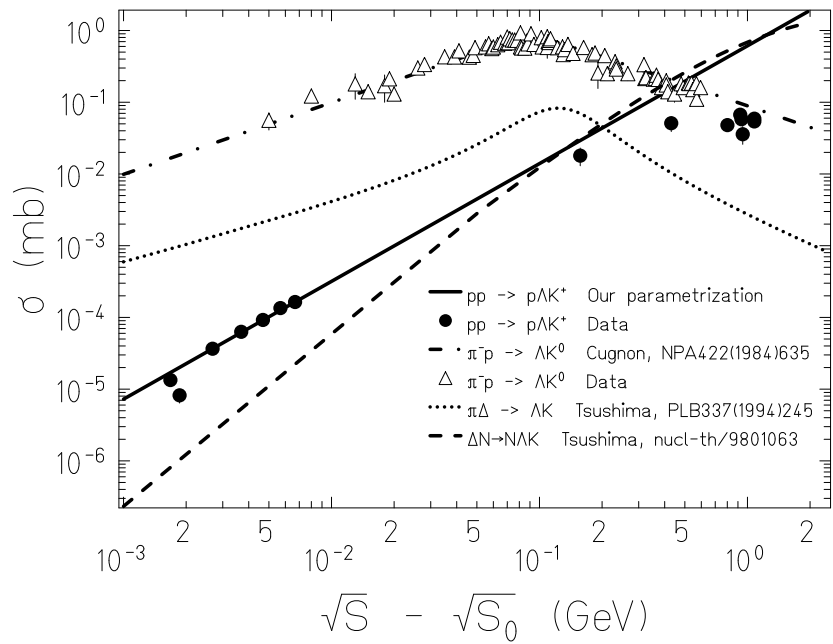}
$$
\caption{\textit{Kaon production cross sections as compared with the available
experimental data.}}
\label{spsouq}
\end{figure}

\begin{figure}[h]
\epsfxsize=11.cm
$$
\epsfbox{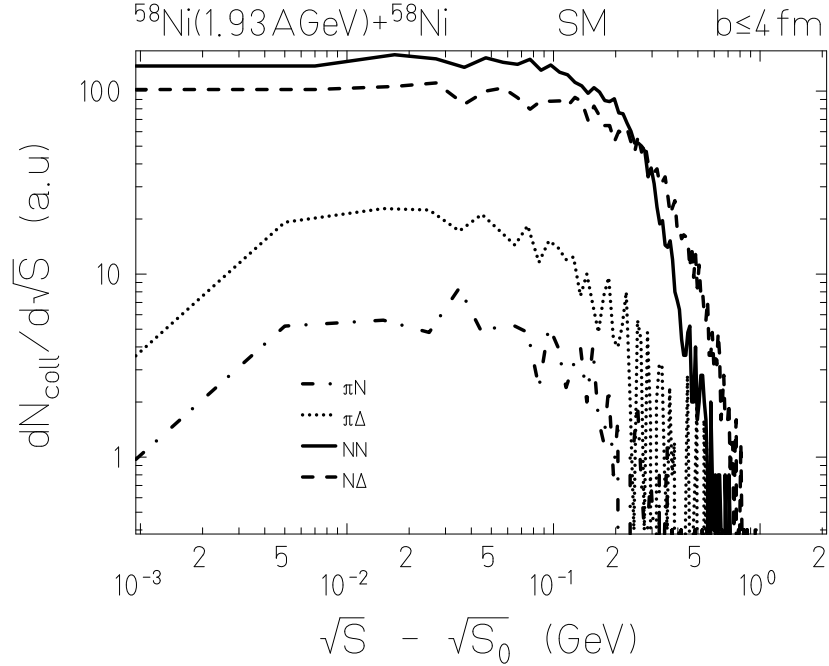}
$$
\caption{\textit{Collision distribution producing a kaon.}}
\label{spsouq}
\end{figure}

\begin{equation} 
 u_{opt}^K = \sqrt{(\vec{k}-g_v\vec\Sigma_v)^2 + m_K^2 + m_Kg_s\Sigma_s }+
  g_v\Sigma_v^0     - \sqrt{k^2 + m_K^2 }
\end{equation} 
 and rescatter with the nucleons. The cross section of the rescattering
 of the produced kaons is a parametrisation of the results presented in 
 \cite{dowa}. 

\section{Kaon's: messenger from the high density zone?}   

Naively one expects that K's are created in the high density zone of the
reaction. Because we are at subthreshold energies the kaon is produced easiest
if there is a nuclear resonance in the entrance channel, which - due to
its higher mass as compared to a nucleon - needs only a smaller relative
momentum with respect to its scattering partner to overcome the threshold.
The shorter its mean free path, the more probable the resonance encounters
a nucleon before it disintegrates. Because $\lambda \propto {1 \over \sigma
\rho}$  the kaon production via resonances appears preferably at high densities.
When created, the kaon is still surrounded by nuclear matter and may encounter
a collision with nucleons while the system disintegrates. The cross section 
for rescattering of
$\approx 12 mb$ is small but nonnegligable. Such a collisions may destroy
the information about the high density zone which is carried by the K's.

Fig 2. displays the distribution of nuclear densities at the places where
the K's are produced. We see that even for a system as small as Ni+Ni 80\%
of the K's are produced at a density above normal nuclear matter density 
and 60\% do not rescatter at a density lower than normal nuclear matter
density. Hence K's can really be regarded as messenger from the high
compressed phase of the reaction. It is the only meson which has this
property at the energy considered in this article.

\begin{figure}[h]
\epsfxsize=11.cm
$$
\epsfbox{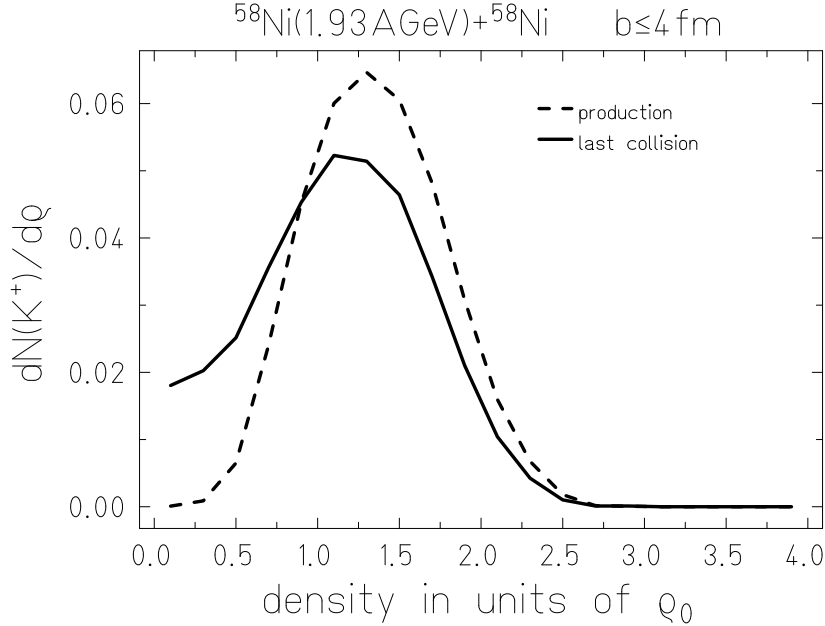}
$$
\caption{\textit{Density distribution of the kaon production and of the last
collisions of the kaons with the surrounding nucleons.}}
\label{spsouq}
\end{figure}

\section{ In-plane flow of the K's}
Recently it has been speculated \cite{ko7,ko} that the in-plane flow of the K's is
directly related to the relative strength of the scalar and vector potential
of K's in nuclear matter. Therefore it is interesting to discuss in detail
the origin of the in-plane flow of the K's, which is much lower than
the in-plane flow of the nucleons measured in the same reaction.

\begin{figure}[h]
\epsfxsize=11.cm
$$
\epsfbox{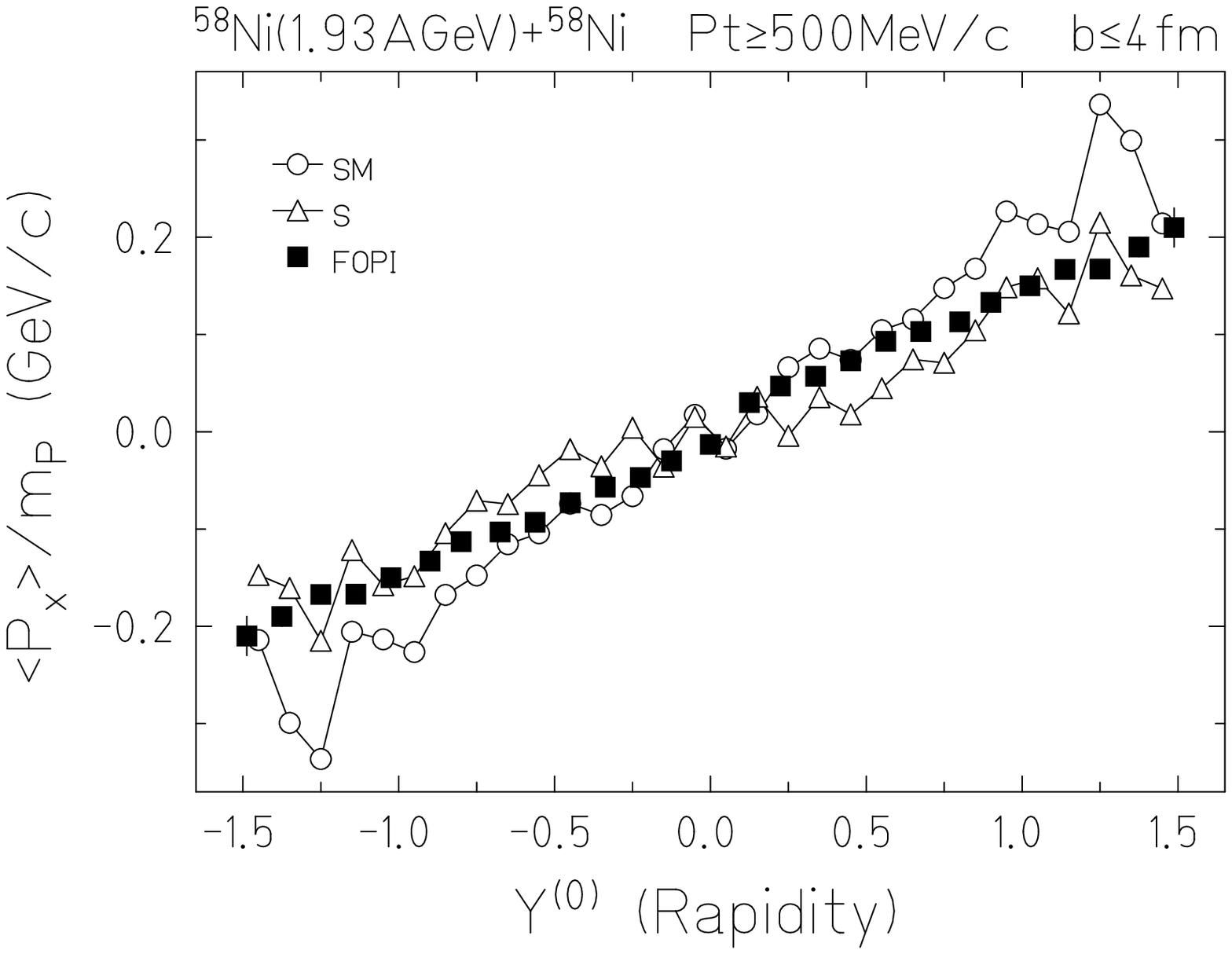}
$$
\caption[]{\textit{Experimental nucleon flow \cite{rit} as compared with filtered 
QMD simulations for 2 different equations of state, a soft equation of state (S)
and a soft momentum dependent equation of state (SM).
}}
\label{spsouqs}
\end{figure}

\begin{figure}[h]
\epsfxsize=11.cm
$$
\epsfbox{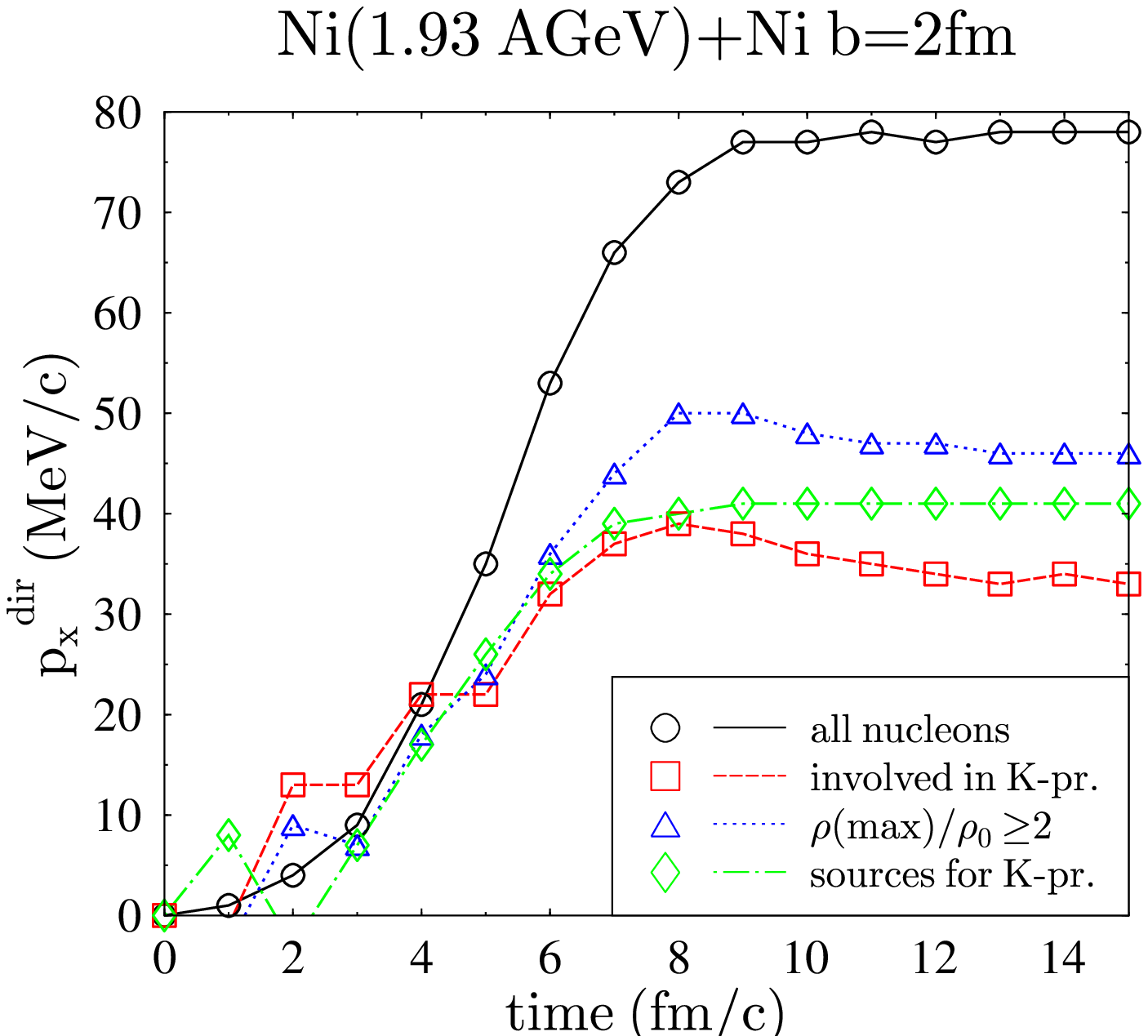}
$$
\caption{\textit{In-plane flow of different classes of nucleons.}}
\label{spsouq}
\end{figure}

Before discussing the kaon flow it is, however, necessary to verify that the
simulations reproduce the nucleon flow although, as we will see later, both
are only very loosely connected. The experimental nucleon in-plane flow as compared to
the simulation is displayed in fig. \ref{spsouqs}. We see that the soft
momentum dependent interaction \cite{aic} reproduces the experimental flow as well as
the less realistic static soft equation of state.

Fig.5 displays the time evolution of the in-plane flow of different classes 
of protons. The average directed in-plane flow is defined by
\begin{equation}
p_x^{dir} =  {1 \over N}\sum_{i=1}^N p_x^{i} \cdot sign 
(y_{cm}^{i})
\end{equation}
where $y_{cm}^i$ is the rapidity of the nucleon
i in the nucleus-nucleus center of mass system and $p_x^i$ is the momentum 
of the nucleon i in the
direction of the impact parameter. The calculation has been done at an impact  
parameter of 2 fm. 

The averaged directed in-plane flow of all nucleons is given by a line. 
We see that it reaches asymptotically 80 MeV/c. For the  dotted line we have counted only
those nucleons which have passed a density of $\rho / \rho_0 \geq 2$.
As already found in ref. \cite{jae} the in-plane flow of nucleons which have
passed the high density zone is considerably smaller than that of all nucleons.
If we include in the calculation of $p_x^{dir}$ only those nucleons, which
have been involved in a collision, in which a kaon has been produced (dashed
line), we observe about the same in-plane flow as for those nucleons which have
passed the high density zone. This agreement is compatible with the fact that
K's are produced in the high density zone as shown in fig.1. More interesting
than the in-plane flow of the protons is the in-plane flow of the sources in
which the K's are produced according to phase space, i.e. isotropically:
\begin{equation}
p_x^{dir} =  {1 \over M}\sum_{i=1}^M (p_x^{1i}+p_x^{2i}) \cdot sign 
(y_{cm}^{1i}+y_{cm}^{2i}).
\end{equation}
Here the sum runs over all collisions in which a kaon is produced. 1 and
2, respectively, mark the two hadrons which scatter in these collisions. The
result is displayed as the dashed-dotted line. This in-plane flow has
about the same size as that for the nucleons involved in a kaon  producing
collision and is not twice as large as one may think naively. Hence
the in-plane directed source velocity $p_x^{dir}$ / mass is about half
as large as the directed in-plane velocity of the participating
nucleons.

\begin{figure}[h]
\epsfxsize=8.5cm
$
\epsfbox{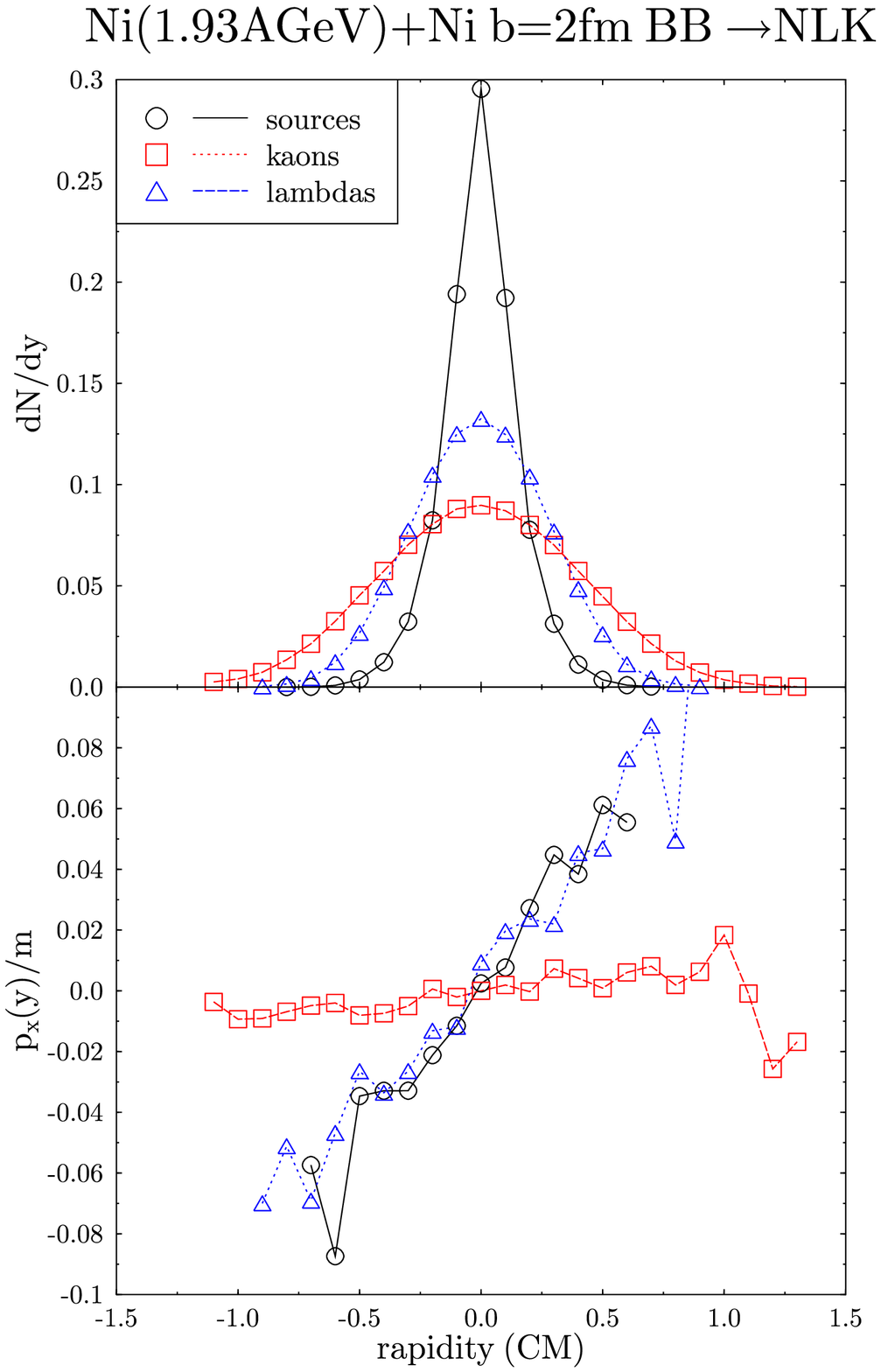}
$\hspace*{-10mm}
\epsfxsize=8.5cm
$
\epsfbox{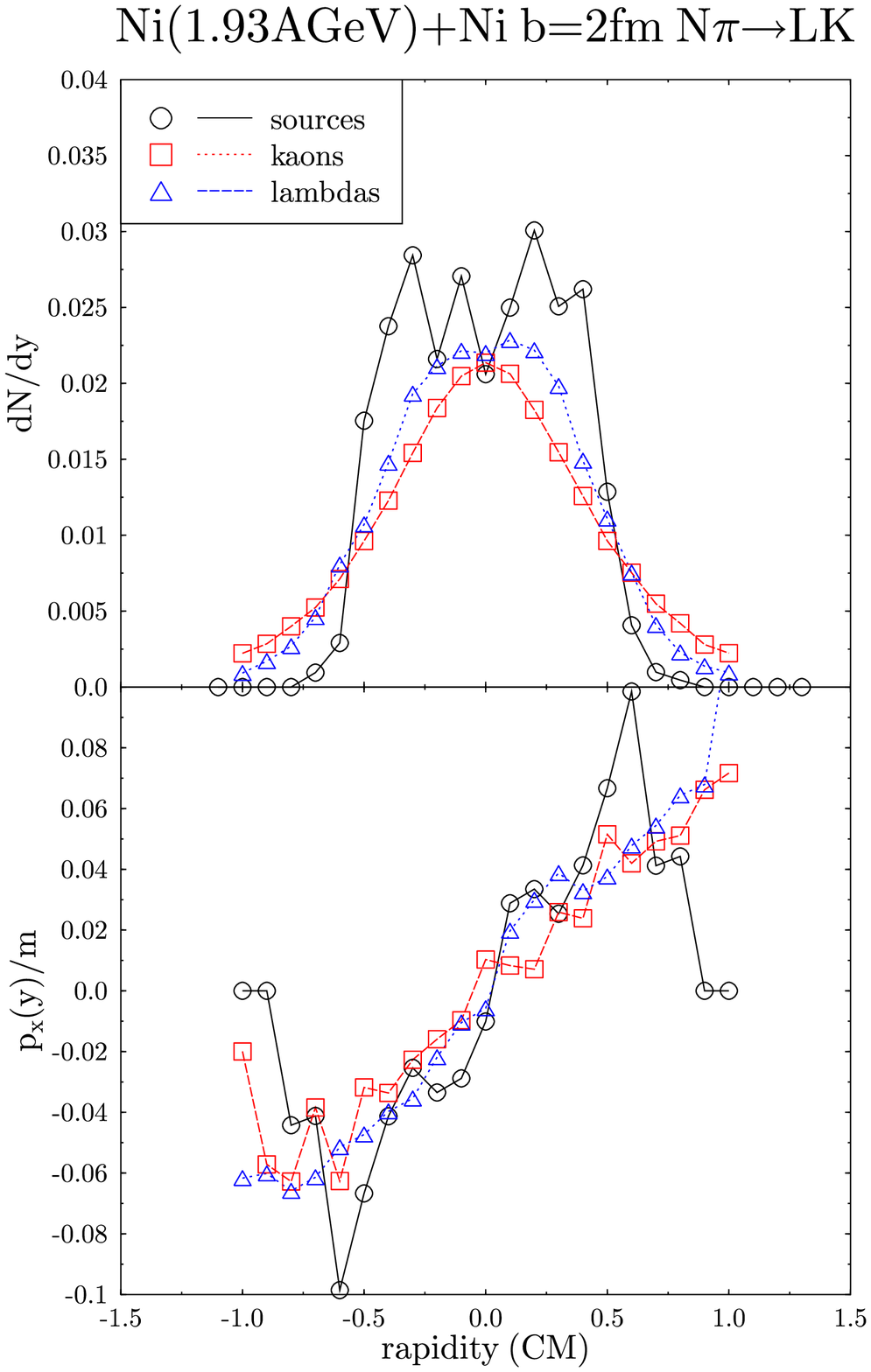}
$ \hspace*{-10mm}
\caption{\textit{Rapidity distribution (top) and in-plane velocity (bottom)
 of K's, $\Lambda$'s and the
sources for the reaction baryon+baryon $\rightarrow \Lambda K N$ (left) and
$\pi N \rightarrow K \Lambda$ (right). }}
\label{spsouq}
\end{figure}

The reason for this astonishing result is displayed on the left hand side
of fig. 6 where we show
the rapidity distribution of the kaon producing sources as well as that of the
$\Lambda 's$ and of the K's at the moment of their creation (i.e. before
rescattering) in the reaction baryon+baryon $\rightarrow \Lambda K N$. 
We see that - due to kinematical reasons - most of the sources
are centered very close around midrapidity in the nucleus-nucleus center of
mass system. Hence it happens quite often that one of the reaction partners
has a negative and the other a positive rapidity. In this case the average
$p_x$ of the both nucleons points into opposite directions and the vector
sum becomes small. This explains the low
value of the in-plane flow of the source velocity.
The three body phase space decay $\Lambda K N$ broadens the rapidity
distribution of the sources considerably. This lowers the in-plane flow of the K's a second
time. At a given  rapidity we find K's from sources with quite different
rapidities and consequently with quite different in-plane source
velocities.  Hence the in-plane velocity of those K's is an average 
value of the in-plane velocities of the different sources. Because the in-plane
velocity changes sign at mid rapidity, the average value is very small. This
is seen in fig. 6 (left) where the rapidity distribution and the 
directed in-plane velocity  of sources, $\Lambda $'s
and K's are displayed. Fig. 6 (right) displays the same quantities for
the channel $\pi N \rightarrow \Lambda K$. Here the in-plane velocity 
of the sources is of about the same size as that for the 3-body channel.
However, having only two particles in the exit channel, this reaction 
does not spread
the in-plane velocity as far in rapidity as a three body decay. Hence finally
K's coming from a $\pi N$ collision have a larger in-plane flow than those
from a baryon baryon collision.

In conclusion we see that the small value of the in-plane flow of the K's as
compared to that of the nucleons is expected even
if one neglects completely any potential or collisional interaction of the
K's with the nuclear environment. It is caused by two processes: nucleons
which pass the high density zone of the reaction have a much smaller average
flow than the average nucleon and the production of K's according to phase 
space transports the K's far away in rapidity space with respect to the
rapidity of the center of mass of the collision in which it is produced. Thus
the in-plane velocity is smeared out. 

\section{The role of the lifetime of the $\Delta$ resonance}
Most of the collisions in which a kaon is produced involve either directly a
$\Delta$ or the $\pi$ produced by its disintegration. The relative fraction
depends on the lifetime of the $\Delta$. For a long time it has been thought
that the life time of a $\Delta$ with a given mass $m_{\Delta}$ depends on the
phase space for decay by.
\begin{equation}
\tau = {1\over \Gamma};\quad  \Gamma \propto \int  d^3p_\pi d^3p_N |<i|M|f>|^2
\delta^4(p_i-p_f)   
\end{equation}

\begin{figure}[h]
\epsfxsize=11.cm
$$
\epsfbox{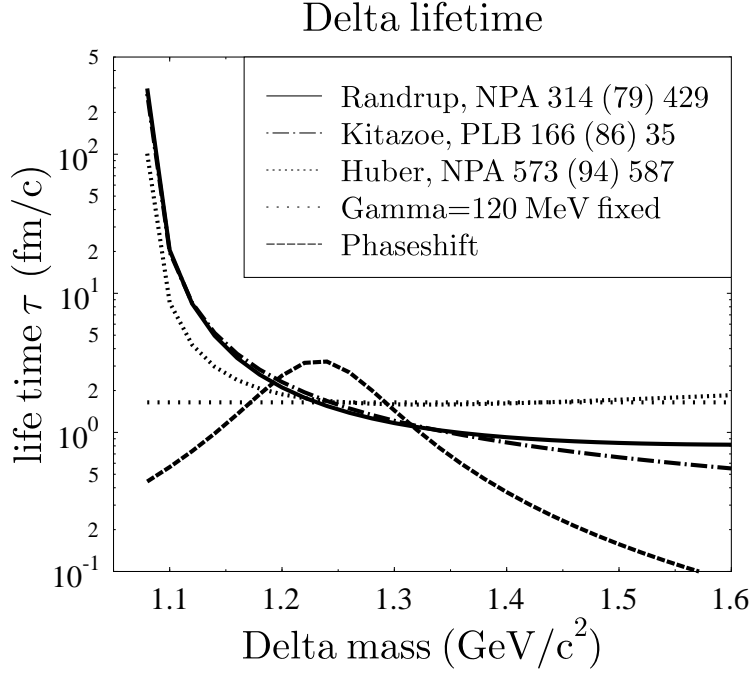}
$$
\caption{\textit{Lifetime of the $\Delta$ resonance as a function of its mass
as used in different simulation programs as compared to the formula of a Breit-Wigner
resonance.}}
\label{spsouq}
\end{figure}

Usually one assumed that the matrix element is independent of the mass of the
resonance and therefore the mass dependence is given by phase space.
Only recently it has been realized \cite{hir} that this formula is only valid
for a well prepared resonance which has the Breit-Wigner distribution of their
mass. In simulations we are confronted with a different situation. The
variance of the energy of the incoming particles is zero in BUU and small as
compared to the variance of the resonance in QMD simulations. Therefore 
the center
of mass energy - which corresponds to the resonance mass - is determined with a
very good precision and its variance is much smaller than the variance of the Breit 
Wigner distribution which characterizes the resonance.    
In this case
the lifetime of  is given by the more
general formula \cite{hir}
\begin{equation}
\tau = {d \delta \over d E} = {\Gamma/2 \over (E-E_0)^2 + \Gamma^2/4}.
\end{equation}
Whereas in the former case (eq.4) the lifetime increases with decreasing 
mass of the
$\Delta$ due to the smaller phase space, in the latter case (eq.5) 
the life time has
its maximum at the center of the resonance and approaches zero at energies
well above or below. Fig.7 compares the lifetimes presently used in the
different numerical programs with that calculated with the above equation.
The influence of the different assumptions on the  $\Delta$ lifetime
on the kaon flow will become increasingly important if one lowers
the beam energy because more and more $\Delta's$ will be produced close to
the threshold due to the limited available energy. At the reaction considered
here the correct equation increases the average lifetime of the $\Delta$'s 
as compared to the former approaches because due to the sufficient energy  
the $\Delta$'s are created preferably at the center of the resonance.

\begin{figure}[h]
\epsfxsize=11.cm
$$
\epsfbox{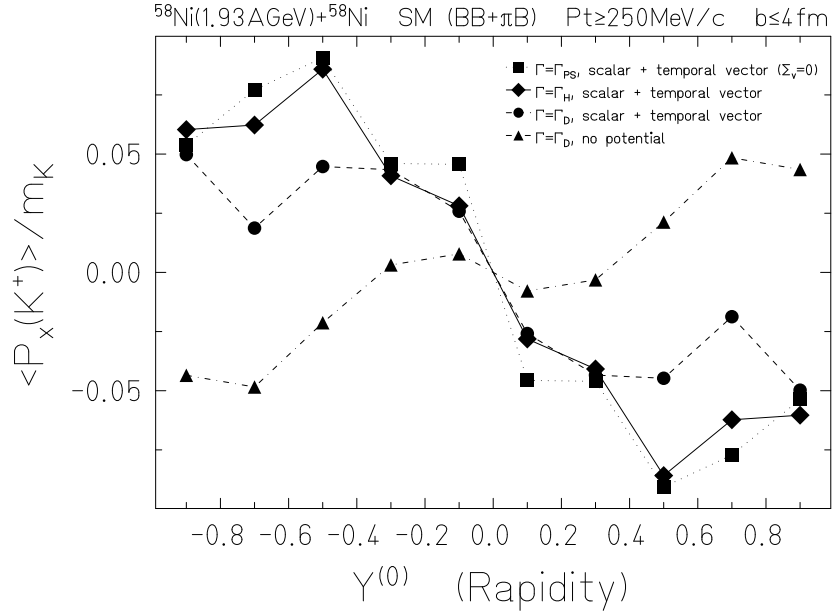}
$$
\caption{\textit{Influence of the lifetime of the $\Delta$ resonance in
central collisions on the in-plane flow of kaons. A soft
momentum dependent equation of state is used and the experimental cut has
been applied.
}}
\label{spsouq}
\end{figure}

\begin{figure}[h]
\epsfxsize=11.cm
$$
\epsfbox{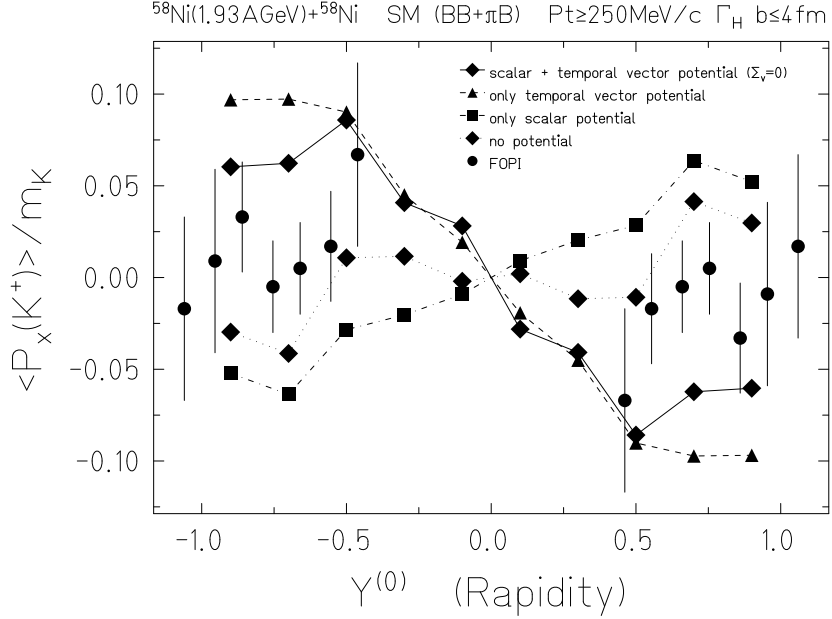}
$$
\caption{\textit{The influence of the different potential parts 
of the K's in 
central collisions on the in-plane flow of kaons. A soft
momentum dependent equation of state is used and the experimental cut has
been applied.}}
\label{spsouq}
\end{figure}

Fig.8 displays the influence of the different lifetimes of the $\Delta$
resonance on the kaon in-plane flow. We compare the flow calculated by using the 
$\Delta$-lifetime as derived by Huber et al. \cite{hu} ($\Gamma_H$) with that obtained 
by eq.6 ($\Gamma_{PS}$) and that obtained by the parameterization
of $\tau = {d \delta \over d E}$ ($\Gamma_D$) from data of \cite{loc}.
For these calculations we have set $\overrightarrow \Sigma_v=0$. We see that
the differences for the in-plane flow are considerable what offers the
possibility to study the $\Delta$-lifetime in medium.

How our calculation compares with experiment is seen in fig.9 where we display
the data of the FOPI collaboration \cite{fopi} in comparison with different 
calculations. For this purpose we have filtered our data. The limitation
to particles with a transverse momentum larger than 250 MeV/c decreases the
statistics by a large factor.  For $\Gamma_H$ we display calculations in
which we set $\Sigma_s$ and $\Sigma^0_V$, respectively, to zero   and compare
them with the result obtained by including and excluding the total static 
potential interaction. A more realistic
$\Delta$-lifetime as compared to \cite{ko} and the addition of the momentum dependent part of the vector
potential balance each other at that energy. 
Thus finally we come close to the results of Ko \cite{ko}. Finally, we 
display in fig. 10 the influence
of the vector part of the vector potential on the in-plane flow of kaons. The temporal
vector potential represents the case when we take only the term $\Sigma^0_V$
of
the potential into account (i.e. $\overrightarrow \Sigma_v=0$). The additional term 
$\overrightarrow \Sigma_v$ increases
the flow of kaons. This is easy to
understand: We have seen (fig.3) that all kaons come from the high density
fireball region where due to the symmetry of the system the momentum
distribution of the baryons is symmetric with respect to the nucleus-nucleus 
center of mass system. Assuming that $\Sigma_v^\mu$ is proportional 
to the baryon current $j^\mu$ the scalar product in the optical potential,
$\vec k \cdot \vec \Sigma_v$, is zero.  Therefore for symmetric systems
the optical potential is reduced to  

\begin{equation} 
 u_{opt}^K = \sqrt{{k}^2+(g_v\vec\Sigma_
v)^2 + m_K^2 + m_Kg_s\Sigma_s }+ g_v\Sigma_v^0     - \sqrt{k^2 + m_K^2 }
\end{equation} 

The $\vec\Sigma_v^2$ term has the same sign as $\Sigma_v^0$ and 
depends in the same way 
on the local baryon density. It enhances therefore the action of 
$\Sigma_v^0$ term as seen in fig. 10.  We would like to mention that in ref.
\cite{fu} an opposite result has been found.

\begin{figure}[h]
\epsfxsize=11.cm
$$
\epsfbox{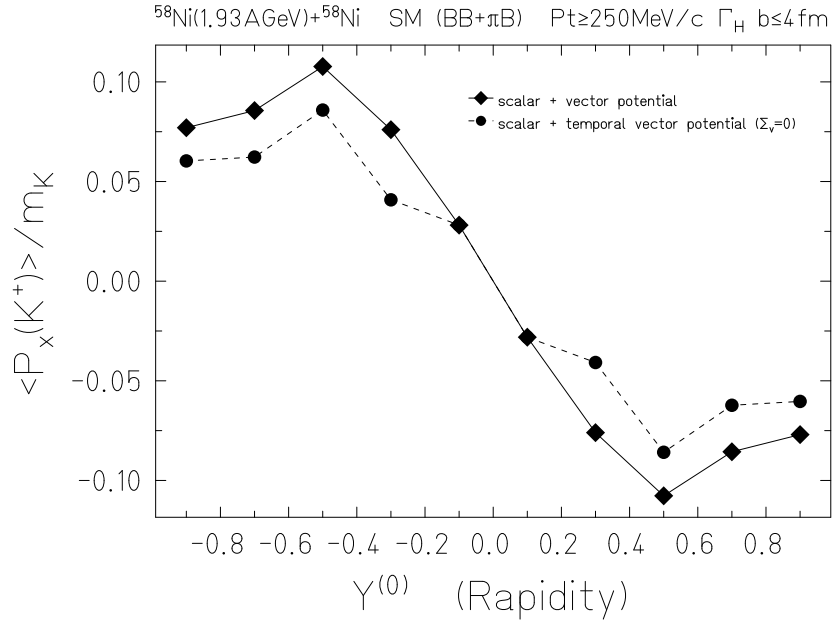}
$$
\caption{\textit{The influence of the different vector potential parts 
of the K's in 
central collisions on the in-plane flow of kaons. A soft
momentum dependent equation of state is used and the experimental cut has
been applied.}}
\end{figure}

In conclusion we have found that the low value of the in-plane flow of the
kaons as compared to nucleons has little to do with the interaction of the
kaons after their creation. It is a consequence of the kinematics before
and at the time point of the creation. The potential interaction of the kaons 
with the nuclear environment as well as the life time of the nuclear resonances and
the rescattering modifies the directed in-plane kaon momentum. In addition, 
the value obtained in the simulation is modified by the experimental cuts.

Therefore it is premature to interpret the kaon flow as a signal for strength of
one or the other of these processes. Because each of the above mentioned
processes is interesting in itself exciting physics lie ahead of us on the
way to interpret the data on the kaon flow.


\begin{thebibliography}{99}
\bibitem{aik} J. Aichelin and C.M. Ko Phys. Rev. Lett. 55 (1985) 2661
\bibitem{har} C.Hartnack et al., Nucl. Phys. A580 (1994) 643
\bibitem{ko7} G.Q. Li, C.M. Ko and B.A. Li, Phys. Rev. Lett 74. (1995) 235 
\bibitem{aic} 
J. Aichelin, Phys. Rep. {\bf 202}, 233 (1991), and references therein.
\bibitem{scha} J. Schaffner et al, Phys. Lett B334 (1994) 268 
J. Schaffner-Bielich et al., Nucl. Phys. {\bf A625}
(1997) 325
\bibitem{dowa} 
C.B. Dover and G.E. Walker, Phys. Rep. {\bf 89}, 1 (1982)
\bibitem{ko}G.Q. Li et al., Phys. Rev {\bf C57} (1998) 434
\bibitem{jae}C. Hartnack, H. St\"ocker and W. Greiner, Proc. on the NASI on the
Nuclear Equation of State, Peniscola (Spain), NASI Series B, Vol. 216A,
Plenum, New York, p. 239. \\
J. J\"anicke and J. Aichelin,
Nucl. Phys. {\bf A 547} (1992) 542
\bibitem{rit} J. Ritman and al., Z. Phys. {\bf A352} (1995) 355 
\bibitem{hir} 
J. Aichelin, Proceedings of Hirschegg 1997, ed. 
 by H. Feldmeier and W. N\"orenberg, and references therein.
P. Danielewisz, Proceedings of Hirschegg 1997, ed. 
 by H. Feldmeier and W. N\"orenberg, and references therein.
\bibitem{hu} S. Huber and J. Aichelin, Nucl. Phys. {\bf A573} (1994) 587
\bibitem{loc} W. Lock and D. Measday, Intermediate energy nuclear physics,
ed. Nederlandse Boekdruk Industrie n.v., 1970.
\bibitem{fopi} D.Best et al., Nucl. Phys. {\bf A625} (1997) 307
\bibitem{fu} C. Fuchs, nucl-th 980148

\end{thebibliography}
\end{document}